\documentclass[epj]{svjour}

\newcommand{\lsim}{\buildrel < \over {_\sim}}

\def\sinhat{\hat{s}^2}
\usepackage{graphics}
\begin{document}
\title{Future Directions in Parity Violation}
\subtitle{From quarks to the cosmos}
\author{Michael J. Ramsey-Musolf\inst{1,\, }\inst{2}
}                     
%
%
\institute{California Institute of Technology, Pasadena, CA 91125 USA \and University of Wisconsin-Madison, Madison, WI 53706 USA}
\date{Received: date / Revised version: date}
%
\abstract{
I discuss the prospects for future studies of parity-violating (PV) interactions at low energies and the insights they might provide about open questions in the Standard Model as well as physics that lies beyond it. I cover four types of parity-violating observables: PV electron scattering; PV hadronic interactions; PV correlations in weak decays; and searches for the permanent electric dipole moments of quantum systems. 
} 
\maketitle
\section{Introduction}
\label{intro}
This year we mark the fiftieth anniversary of the discovery of parity violation (PV) in the $\beta$-decay of polarized $^{60}$Co\cite{Wu:1957my} and the decay of polarized muons\cite{Garwin:1957hc}, confirming the prediction of Lee and Yang\cite{Lee:1956qn}. The subsequent history of the field is remarkably rich and diverse, as studies of PV interactions have provided key insights about both the nature of the fundamental weak interaction of leptons and quarks as well  the internal structure of strongly interacting systems such as the proton and nuclei. As the interest in this series of PAVI meetings amply demonstrates, the study of PV effects remains an engaging topic that will undoubtedly demand our efforts for many years to come. In this talk, I will give my perspective on how the field may develop over the coming decade, bearing in mind that there are likely to be surprises and new ideas that we cannot foresee at present. In doing so, I will endeavor to give a representative reference to the relevant literature (without providing a full-fledged review), and I apologize in advance to anyone whose work I inadvertently omitted.  

In thinking about PV, I like to put the field in the broader context of fundamental symmetries and their relation to the history of the cosmos. Broadly speaking, I break the particle physics of the universe into three eras: (1) the present era, during which the broken symmetries of the Standard Model (SM) provide a remarkably successful framework for explaining a host of microscopic and astrophysical phenomena; (2) the era running from the Big Bang to the moment when electroweak symmetry was broken; and (3) the brief period of electroweak symmetry breaking (EWSB). Remarkably, studies of PV interactions at low energies has something to say about the microphysics of each of these eras, as I hope to convince you below. More extensive discussions of PV in the  context of low-energy weak interaction studies can be found in Refs.~\cite{Erler:2004cx,Ramsey-Musolf:2006ur,Ramsey-Musolf:2006ik,Ramsey-Musolf:2006vr}.

\section{PV: the Standard Model Era}
\label{sec:SM}

It would take an entire talk to summarize the successes of the SM, so I will focus on what I call \lq\lq unfinished business" of the theory: the simplest part of the SM to write down is the hardest to solve. Explaining how QCD gives rise to the nonperturbative structure of hadrons and nuclei and governs their interactions is this unfinished business, and it remains a central quest for nuclear physicists. In the past decade, PV electron scattering (PVES) has proven to be an important tool in pursuing this quest. Nearly two decades ago, polarized deep inelastic scattering started to create a \lq\lq spin crisis" suggested that the naive quark model picture of the proton was flawed, as the quarks carried perhaps $\sim 30\%$ of the proton's spin (for a review, see {\em e.g.} Ref.~\cite{Filippone:2001ux}). There were also indications that strange quarks, which do not appear in the quark model picture, contributed as much as 10\% of the proton spin and were polarized anti-parallel to it. At the same time, people began to wonder whether strange quarks could also play an important role in the nucleon's electromagnetic structure, and early computations by Jaffe\cite{Jaffe:1989mj} and others suggested that a large strange quark contribution was not unreasonable (for recent reviews of the theoretical literature, see, {\em e.g.}, Refs.~\cite{Ramsey-Musolf:2005rz,Beise:2004py,Spayde:2003nr,Beck:2001dz}). 

As first suggested by Kaplan and Manohar\cite{Kaplan:1988ku} and followed by with experimental ideas by McKeown\cite{Mckeown:1989ir}, Beck\cite{Beck:1989tg} and others, PVES became the probe of choice to look for these strange quark contributions. As a result of the efforts of the SAMPLE\cite{Spayde:2003nr}, PV A4\cite{Maas:2005au}, HAPPEX\cite{Acha:5}, and G0\cite{Armstrong:2005hs} experiments, we now have limits on the size of the strange quark effects. The results are being summarized by other speakers, but I like to characterize them in terms of the isoscalar form factors as these are the only ones to which strange quarks contribute. In the case of the isoscalar magnetic moment, for example, we no know that strange quarks are responsible for at most $(20\pm 15)\%$.  The strange quark contributions to the nucleon electric form factors are similarly bounded to be a small fraction of the total. At the outset of this program of PV studies, various theoretical expectations suggested the possibility of considerably larger effects. Obtaining these experimental results that now rule out many of those early expectations represents a substantial achievement for the field.

The successful use of PVES to probe nucleon strange-ness has depended on a close interplay of experiment and theory. Obtaining robust extractions of bounds on the strange vector form factors from the measured PV asymmetries required a careful delineation of various contributions and sources of theoretical uncertainty \cite{Musolf:1993tb}, including hadronic contributions to electorweak radiative corrections \cite{Zhu:2000gn,Musolf:1990ts}, nuclear wavefunction and many-body effects \cite{Musolf:1994gr,Hadjmichael:1991be,Schiavilla:2002uc,Liu:2002bq}, and isospin breaking in the nucleon \cite{Miller:1997ya,Kubis:2006cy}. Interpreting the results in QCD has inspired recent work using lattice QCD methods\cite{Dong:1997xr,Lewis:2002ix,Leinweber:2004tc,Leinweber:2006ug}. Calculating the strange quark effects requires evaluation of \lq\lq disconnected" operator insertions on the lattice -- a highly non-trivial technical challenge. Given the tight bounds on the strange form factors, obtaining robust results from direct lattice computations stands as an important future milestone for the lattice community.

\subsection{PV DIS: Beyond the parton model?}

In essence, PVES gave us a clear look into the low-energy \lq\lq internal landscape" of the proton beyond the quark model picture. In the future, I think it might play the same role as a way to look beyond the parton model description that has been enormously successful in explaining deep inelastic and Drell-Yan processes. As discussed in detail at this meeting, there exists considerable interest now in revisiting the PV deep inelastic scattering (DIS) experiment that was carried out at SLAC in the 1970s  and that provided the smoking gun for the SM picture of the weak neutral current interaction\cite{SLAC}. In addition to using a more precise version of that experiment to test the electroweak part of the SM and look for new physics, it could also shed new light on the structure of the proton. 

One aspect that particularly intrigues me is the possibility to look for higher twist (HT) effects. Such effects generate corrections to parton model expectations that go as powers of $\Lambda^2_{\rm had}/Q^2$ and that are highly suppressed at the energies where parton distribution functions are typically determined. At lower energies, however, their effects should become more apparent. In this respect, there exists a puzzle emerging from inclusive, inelastic electroproduction data from Jefferson Lab. As detailed extensively in Ref.~\cite{Melnitchouk:2005zr}, the expected HT contributions to structure functions do not behave as expected. Plotting $2x_BF_1(x_B, Q^2)$ vs $x$, for example, one sees that the data fluctuates  about the curves derived from parton distribution functions, that characterize the leading twist contribution. The lower the $Q^2$, the larger the fluctuations. One can clearly see as well that the location of the peaks corresponds to various well-known resonances, indicating departures from incoherent scattering from uncorrelated partons. One can then convert the structure function data into moments and plot the moments as a function of $Q^2$. After subtracting out the known elastic contribution, one finds remarkable agreement with parton model predictions for the lowest moments. 

The puzzle is: What happened to the parton correlations that are naively responsible for the resonance structure in the structure functions and that ought to show up as important HT corrections to the  moments? To be more concrete, consider the momentum sum rules that relate the structure function moments to matrix elements of local operators:
\begin{eqnarray}
\label{eq:momsum}
M_j^{(n)}(Q^2) &= &\int\, dx_B\, x^{n-j}_B\, F_j(x_B,Q^2) \\
\nonumber
&\propto& \sum_{n,j}{\tilde C}_{j,k}^{(n)}(Q^2,\mu^2,g)\, A_k^{n}(\mu)
\end{eqnarray}
where $j$ labels the structure function; the $A_k^{(n)}(\mu)$ are forward matrix elements operators ${\cal O}^{(n)}_k$ having spin $n$ and being labeled by index \lq\lq $k$"; $\mu$ is the renormalization scale; and the ${\tilde C}_{j,k}^{(n)}$ are Wilson coefficients whose $Q^2$-dependence can be computed in perturbation theory. The twist of an operator is defined as its dimension $d_{\cal O}$ minus its spin: $\tau=d_{\cal O}-n$. For twist two operators, the Wilson coefficients carry only a logarithmic dependence on $Q^2$ (summed to all orders using the renormalization group). For twist four operators, the ${\tilde C}$ carry an additional $1/Q^2$ power dependence. The corresponding operator matrix element introduces an additional $\Lambda^2_{\rm had}$ in the numerator (with a coefficient determined by the details of the operator matrix element), leading to the $\Lambda^2_{\rm had}/Q^2$ power correction to the twist two logarithmic $Q^2$-dependence. Successively higher orders in the twist expansion imply higher order power corrections of this type. Generically, one expects that for $Q^2\sim 1$ GeV$^2$, these power corrections should become quite important. The puzzle is that in the low-$Q^2$ data for the lowest moments, they do not seem to be there in a significant way. 

It is interesting to note that the$\tau=4$ operators generally probe correlations between quarks and gluons in the nucleon. Illustrative examples include\cite{Jaffe:1982pm}
\begin{eqnarray}
\nonumber
\Delta\cdot Q_n^{1(k,\ell)} & = & g\,{\bar q}_R\ {\not\!\!\Delta}\, {\overleftarrow d}^\ell\, {\overrightarrow d}^k q_R\,
{\bar q}_R\ {\not\!\!\Delta}\, {\overrightarrow d}^{n-2-k-\ell}q_R\\
\label{eq:twistfour}
\Delta\cdot Q_n^{8(k)} & = & i {\bar q} {\overleftarrow d}^k\Delta_\alpha\gamma_\beta G^{\alpha\beta}\, {\overrightarrow d}^{n-1-k} q
\end{eqnarray}
where $\Delta_\mu$ is  a light-like vector contracted with all free operator indices, $G^{\alpha\beta}$ is the gluon field strength tensor, and $d=i\Delta\cdot D$ with $D_\mu$ being the gauge covariant derivative. From the examples in Eq.~(\ref{eq:twistfour}) one can see that twist four operators probe quark-quark correlations ($\Delta\cdot Q_n^{1(k,\ell)}$) or quark-gluon correlations ($\Delta\cdot Q_n^{8(k)}$). Thus, probing the effects of such operators is a way to look beyond the parton model picture of the nucleon in which scattering occurs incoherently from quasifree (uncorrelated) constituents. 

Simple curve fitting to power corrections suggests that contributions from a given order in the twist expansion beyond twist two may be large, but that cancellations occur between successive orders (see, {\em e.g.}, Ref.~\cite{Osipenko:2003bu}). This may be plausible, but is not yet convincing, since we do not know the non-power law $Q^2$-evolution  of HT contributions to the structure functions ({\em e.g.}, the logarithmic dependence of the ${\tilde C}$s). It seems equally possible that the leading twist/parton model description fits the lowest moments because the matrix elements of the relevant HT operators are suppressed rather than because of a conspiracy of cancellations.  Ideally, one would like to have access to a different combination of HT operators than those entering parity conserving scattering to help sort out among these possibilities.  In this sense, PV DIS may provide us with a novel tool for looking beyond the parton model  description of the nucleon that has worked so well for inelastic electroproduction, just as PVES allowed us to look beyond the quark model (for strange quarks) description of the elastic EM nucleon properties. 

To carry the parallel an bit further,  theoretical computations of the nucleon strange magnetic moment ($\mu_s$) based on dispersion theory suggested that the portion of $\mu_s$ arising from the kaon cloud (resonating into the $\phi$) could be comparable in magnitude to the isoscalar magnetic moment, but that its effects in  electromagnetic interactions are hidden by a conspiracy of cancellations\cite{Jaffe:1989mj,Hammer:1999uf}. Without looking directly for $\mu_s$ using PVES, we would never have been able to test this scenario. The results from the PVES experiments now imply that the picture is more complicated and that it will take additional theoretical efforts to understand the dynamics of $s{\bar s}$ pairs in the vector channel. I suspect that we may be in a similar situation with HT effects and that PV DIS could provide important information that complements what we know from parity conserving electroproduction.

In exploring this possibility, I believe there is considerable theoretical work to be completed. The experimentalists are far ahead of theorists in developing the PV DIS program, but I am confident that we can lay out the framework for a systematic and interpretable program just as we did over a decade ago for elastic and quasielastic PVES. In doing so, we should address several questions: What is the non-power law $Q^2$-dependence of the twist four Wilson coefficients? What information on HT operators from parity-conserving electroproduction can be used as input for the PV case? Which twist four operators are the PV asymmetries most sensitive to and which kind of quark-quark or quark-gluon correlations do they probe? What would we expect for the corresponding matrix elements in QCD? I hope that by the time of the next PAVI meeting in two years we will begin to have some answers to these questions. 

\subsection{Hadronic PV: four quark operators in the nuclear domain}

The purely hadronic weak interaction (HWI) has been a topic of considerable scrutiny over the years, and yet it remains enigmatic. What makes the HWI both interesting and challenging is that it involves a complex interplay of weak and strong interactions. The effects of perturbative QCD can be computed fairly reliably, as they lead to the running of four quark operators from the electroweak scale down to hadronic scales. The real difficulties arise in understanding the low-energy matrix elements of those operators. Short of first principles, lattice QCD computations, theorists have relied on a number of methods over the years to try and compute these matrix elements: chiral perturbation theory, quark models, large N$_C$ symmetry, SU(3) symmetry, etc. Unfortunately, the data brings all these efforts up short.

As recently discussed in Ref.~\cite{Ramsey-Musolf:2006dz}, the strangeness changing ($\Delta S=1$) sector presents a number of puzzles that remain to be explained. The reason for the $\Delta I=1/2$ rules is, perhaps, the most well-known. In addition, the non-leptonic decays of hyperons have defied a description within the framework of chiral perturbation theory (ChPT), which has been very successful in describing strong and electromagnetic interactions of hadrons. In the case of the $\Delta S=1$ HWI, one can adequately characterize either the $S$- or $P$-wave non-leptonic decays with ChPT, but not both. As suggested in Refs.~\cite{Borasoy:1999md,Borasoy:1999nt}, one may need to include a host of baryon resonances to arrive at a satisfying description. An explanation of the PV radiative decays such as ${\vec\Sigma}^+\to p\gamma$ is even more elusive. In this case, one measures a PV asymmetry $\alpha_{BB'}$ of the outgoing photon relative to the polarization of the decaying hyperon. This asymmetry arises from an interference of the parity conserving M1 and parity-violating E1 amplitudes. In the pure SU(3) limit, the latter must vanish\cite{Hara:1963gw}. Thus, one would expect $\alpha_{BB'}\sim m_s/\Lambda_{\rm had}$, characteristic of the size of SU(3)-breaking. Experimentally, however, one finds asymmetries four to five times larger magnitude. 

Apparently, the symmetries of low-energy QCD are not terribly helpful when applied to the $\Delta S=1$ HWI. What we don't know is whether this failure reflects the presence of the strange quark, with its problematic mass (not too large, not too small), or more fundamental dynamics associated with four quark matrix elements in hadrons. One way to find out is to study the $\Delta S=0$ HWI, for which the effects of the strange quark ought to be relatively unimportant. The best way for doing so is to study PV effects in hadronic and nuclear systems. Since one would like to get at the effects of the weak interaction, and since the strong and electromagnetic interactions are many orders of magnitude stronger than the weak interaction in $\Delta S=0$ processes, one must rely on PV observables to filter out the latter from the former. 

Unfortunately, the state of our understanding of PV effects in hadronic and nuclear systems remains rather poor, despite years of dedicated experimental and theoretical effort\cite{Ramsey-Musolf:2006dz}. The problem is two fold. First, in order to see the effects of the PV $\Delta S=0$ HWI, experimentalists had to rely on processes in nuclei, such as PV $\gamma$-decays analogous to the hyperon radiative decays. Due to accidental near degeneracies of opposite parity states in  nuclei such as $^{18}$F, the effects of the PV HWI could be enhanced by several orders of magnitude over the nominal $\lsim 10^{-7}$ scale. While such enhancements made the measurement of PV effects eminently more feasible, the use of nuclei also complicated their interpretation. Second, for many years the theoretical framework used to perform this interpretation relied on a picture of PV meson-exchange interactions, wherein the exchanged meson experienced a PV interaction with one nucleon and a parity-conserving strong interaction with the other. To make such a framework practical, one needed to truncate the tower of exchanged mesons to the lightest few that could saturate all the quantum numbers of the low energy PV NN interaction. What the effective vertices in such a picture mean in terms of the underlying HWI is not clear -- especially when one folds in the approximations  needed to perform computations with this meson-exchange model in nuclei. 

Two new developments have changed the prospects for this field, and I am now optimistic that substantial progress can be achieved in the future. First, experimental advances have opened the way to performing measurements of $\sim 10^{-7}$ PV observables in few-body systems, thereby circumventing the need to contend with nuclear many-body issues. A comprehensive program of few-body PV measurements is now being planned at the Spallation Neutron Source, National Institute of Standards and Technology, and elsewhere that will bring about a sea change in the experimental information we have available to us. Second, the framework for interpreting these measurements has been reformulated using the ideas of effective field theory (EFT)\cite{Zhu:2004vw,Ramsey-Musolf:2006dz}. The use of EFT allows us to systematically classify the terms in the PV NN interaction according to powers of a small momentum or mass, labeled by \lq\lq $Q$". The lowest order interaction -- which occurs at ${\cal O}(Q^{-1})$ -- is purely long range and arises from pion exchange, as in the old meson-exchange picture. There is one low energy constant (LEC) associated with this interaction, $h_\pi^1$ (or $f_\pi$ in past literature). The next set of interactions arise at next-to-next-to leading order (NNLO), or ${\cal O}(Q)$ and comprise three classes of terms: (1) a medium range term in the potential associated with two pion exchange; (2) short range contributions to the PV potential that in the meson exchange picture were parameterized by heavy meson exchange; and (3) a long-range pion-exchange meson exchange current (MEC). The medium range term in the potential introduces no new LECs beyond the $h_\pi^1$ entering the LO potential, but it is a qualitatively new effect that had not been systematically included in the old meson exchange picture. The short range terms introduce five additional LECs that correspond to the five independent $S$-$P$ amplitudes that characterize the PV NN interaction at very low energies ($E << m_\pi$): $\rho_t$, $\lambda_t$, and $\lambda_s^{1,2,3}$. The long range MEC introduces one new constant, ${\bar C}_\pi$, that also does not appear in the older framework. Thus, to ${\cal O}(Q)$, one has seven LECs that must be determined from experiment.

The goal of the program at the SNS, NIST, and other facilities is to determine these seven constants from PV experiments with few-body systems. The advantage of the few-body systems is that one can perform precise, {\em ab initio} few-body computations using the lowest order PV NN potential and extract the LECs without encountering many-body nuclear physics uncertainties. A successful completion of this program therefore requires completion of a comprehensive set of few-body computations. For recent work along these lines, see Refs.~\cite{Ramsey-Musolf:2006dz,Liu:2006dm} . At the same time, one would like to compute the LECs from first principles in QCD using the lattice so that -- at the end of the day -- one can confront the experimental values with theoretical expectations. At present, one has estimates for the LECs based on naive dimensional analysis and on a correspondence with the well known \lq\lq DDH" quark model/SU(6)$_w$ ranges for the PV meson-nucleon vertices in the old framework (for a summary of these estimates, see Ref.~\cite{Ramsey-Musolf:2006dz}]. Clearly, one would like to go beyond these model estimates to {\em bona fide} QCD computations, and to that end, initial work using chiral perturbation theory -- needed for extrapolations of lattice computations to the physical light quark domain -- has been performed for $h_\pi^1$ \cite{Zhu:2000fc,Beane:2002ca}. One physical insight derived from these studies is that the role of quark-quark correlations in the four-quark weak matrix elements that generate $h_\pi^1$ could be more important than implied by the DDH quark model/SU(6)$_w$ computation. The results of such QCD computations, when compared with experimental results, could also help us find out whether the $\Delta S=0$ suffers from the same departures from QCD symmetry-based expectations as one finds in the $\Delta S=1$ sector, or whether the latter are just a reflection of the problematic presence of the strange quark.

An additional follow-up to the few-body hadronic PV program would be to revisit the nuclear PV observables. It would be interesting to test whether the NNLO PV NN interaction, determined by the few-body program, can explain the results of nuclear PV experiments when it is used in many-body computations. If so, the results could have consequences for the interpretation of other nuclear weak interaction observables that are sensitive to the effects of four-quark operators. One such observable of particular interest these days is neutrinoless double beta decay ($0\nu\beta\beta$). It is well-known that if one finds a non-zero signal in $0\nu\beta\beta$, one has smoking gun evidence for the Majorana nature of the neutrino. Ideally, one would also like to use a non-zero rate to determine the absolute scale of neutrino mass, since for light Majorana neutrino exchange, the rate is proportional to the square of the effective mass. It is entirely possible, however, that the exchange of some heavy Majorana particle, such as a neutralino in supersymmetry or a heavy right-handed Majorana neutrino, could contribute to the rate at the same level as the exchange of the light Majorana neutrino. In order to separate the two contributions, one would like to be able to compute the effects of the former, which entail calculating nuclear matrix elements of four quark operators. The corresponding nuclear operators can be classified according to $Q$-counting as with the PV NN interaction. What we don't know is whether or not the leading terms in this classification suffice to explain four-quark matrix elements in nuclei. Using the NNLO PV NN interaction to confront nuclear PV data would provide us with a test. If successful, one would then have some confidence that the lowest order terms in the EFT for $0\nu\beta\beta$ give a reasonable description of heavy particle exchange effects in that process.

One alternate probe of the $\Delta S=0$ HWI that remains interesting involves PVES for the $N\to\Delta$ interaction. In contrast to the situation for elastic PVES, the $N\to\Delta$ asymmetry need not vanish at $Q^2=0$ -- a consequence of Siegert's theorem well know in electromagnetic physics. In the real photon limit, the asymmetry has simple form\cite{Zhu:2001br}
\begin{equation}
A_{PV}^{N\to\Delta} = -2\frac{d_\Delta}{C^V_3}\, \frac{m_N}{\Lambda_{\rm had}}+\cdots
\end{equation}
where $d_\Delta$ is a LEC that characterizes the PV $\gamma N\Delta$ electric dipole interaction. It is the $\Delta S=0$ analog of the $\Delta S=1$ E1 amplitude responsible for the PV hyperon radiative decays discussed above. It appears  realistic that a useful determination of $d_\Delta$ could be performed with either the G0 or Q-Weak apparatus. The results of such measurements could shed additional light on the hyperon radiative decay puzzles. If for example, parity-mixing with nucleon resonances is responsible for the large $\alpha_{BB'}$ discussed above, and if the same kind of dynamics apply to the $\Delta S=0$ HWI, then one would expect $A_{PV}^{N\to\Delta}$ to be quite large -- of order $10^{-6}$. On the other hand, if the dynamics are substantially different in the two cases, then naive dimensional analysis would suggest a considerably smaller $N\to\Delta$ asymmetry. Either way, the results would be interesting, and I think it is important to keep the prospects for such a measurement on the radar screen for the future.

\section{PV:  Beyond the Standard Model}
\label{sec:beyond}

The search for physics beyond the Standard Model (BSM) clearly lies at the forefront of particle physics as well as at the intersection of nuclear physics with particle physics and cosmology. This search is motivated by a number of open questions involving the microphysics of the early universe that the SM cannot solve. These questions include: Why is there more matter than antimatter in the universe? Were all the forces of nature unified into a single force at the end of the big bang, and if so, why don't we see this unification in the running of the SM gauge couplings? What is the origin of neutrino mass? Why is electric charge quantized? Why is the electroweak scale so low or, in terms of more familiar quantities, why is the Fermi constant so large? 

Looking ahead to the start of the LHC era, we are hoping to uncover direct evidence for one or more of the possible extensions of the SM that would address some of these puzzles, including supersymmetry, extended gauge symmetries and the associated gauge bosons, extra dimensions and the corresponding modes that would appear in four dimensional spacetime, {\em etc.}. At the same time as we move into the next energy frontier, experiments that push the precision frontier are poised to provide important, complementary information. Most of these experiments involve processes at low-energy and many entail the study of PV observables. I will discuss the future directions in two classes: weak decays and PV electron scattering.

\subsection{Weak decays: PV correlations}

The study of various correlations in weak decays has been a topic of considerable interest for decades, and recent experimental advances are moving the field into a new era. For concreteness,  let me consider the weak decays of the muon and neutron. In the case of polarized muons, the spatial distribution and energy dependence of the outgoing positron or electron are characterized by the Michel parameters that enter the partial rate\cite{michel1,michel2}
\begin{eqnarray}
\nonumber
   d\Gamma& = & {G_\mu^2 m_\mu^5\over 192\pi^3} {d\Omega\over 4\pi} x^2\ dx
  \times \Biggl\{ {1+h(x)\over 1 + 4\eta(m_e/m_\mu)}\\
  \label{eq:michel1}
  &\times&\left[
  12(1-x)+\frac{4}{3}\rho(8x-6)+ 24\frac{m_e}{m_\mu}{(1-x)\over x}\eta\right]\\
  \nonumber
   &\pm&  P_\mu\; \xi\cos\theta \left[ 4 (1-x) + \frac{4}{3}\delta(8x - 6) +
   {\alpha\over 2\pi}{g(x)\over x^2}\right]\Biggl\},
\end{eqnarray}
where $x=|{\vec p}_e|/|{\vec p}_e|_{\rm max}$, 
$\theta=\cos^{-1}({\hat p}_e\cdot{\hat s}_\mu)$, $P_\mu$ is the $\mu^{\pm}$ 
polarization, and $h(x)$ and $g(x)$ are momentum dependent radiative 
corrections. The quantities $\rho$, $\eta$, $\delta$, and $\xi$ are Michel parameters (there are an additional five that do not concern us here). Note that $\xi$ and $\delta$ are associated with the anisotropic part of the charged lepton spectrum that reflects PV in the underlying weak interaction. 
In the SM, one has $\rho=\delta=3/4$, $P_\mu\xi=1$, and $\eta=0$. Deviations from these values would reflect the presence of non $(V-A)\times(V-A)$ interactions. Recently, the TWIST collaboration has completed new determinations of $\rho$, $\delta$, and $P_\mu\xi$, reducing the uncertainty by a factor of two or more over previous determinations\cite{Musser:2004zw,Gaponenko:2004mi,Jamieson:1}. 

The effects of non $(V-A)\times(V-A)$ interactions on these parameters can be described by an effective low-energy Lagrangian
\begin{equation}
\label{eq:leff0}
{\cal L}^{\mu-\rm decay} = -\frac{4 G_\mu}{\sqrt{2}}\, \sum_\gamma \ g^\gamma_{\epsilon\mu}\ 
\ {\bar e}_\epsilon \Gamma^\gamma \nu_e\,  {\bar\nu}_\mu \Gamma_\gamma \mu_\mu
\end{equation}
where the sum runs over Dirac matrices $\Gamma^\gamma= 1$ (S), $\gamma^\alpha$ (V), and $\sigma^{\alpha\beta}$ (T) and the subscripts $\epsilon$ and $\mu$ denote the chirality ($R$,$L$) of the final state lepton and muon, respectively\footnote{The use of the subscript \lq\lq $\mu$" to denote both the chirality of the muon and the flavor of the corresponding neutrino is an unfortunate historical convention.}. The SM gives $g^V_{LL}=1$ with all other $g^\gamma_{\epsilon\mu}=0$. 
For example, deviations of $\delta$ and $P_\mu\xi$ from their SM values are given by
\begin{eqnarray}
\nonumber
\delta-\frac{3}{4} & = & \frac{9}{4}\left[\vert g^V_{RL}\vert^2 + 2\vert g^T_{RL}\vert^2+{\rm Re}\, g^T_{RL}g^{S\, \ast}_{RL} -(L\leftrightarrow R)\right]\\
\label{eq:michel2}
1-\xi\frac{\delta}{\rho} & = & 2 \vert|g^V_{RR}\vert^2+\frac{1}{2}\vert g^S_{RR}\vert^2+\frac{1}{2}\vert g^S_{RR}\vert^2\\
\nonumber
&&+\frac{1}{2}\vert g^S_{LR}-2g^T_{LR}\vert^2\ \ \ .
\end{eqnarray}

Various BSM scenarios can lead to non-zero values for the $g^\gamma_{\epsilon\mu}$ appearing in these PV Michel parameters. For example, a non-zero value of $g^V_{LR,RL}$ can occur in left-right symmetric models when the left-handed (LH) and right-handed (RH) $W$-bosons mix\cite{Herczeg:1996qn}. Similarly, supersymmetric loop graphs can give rise to a non-vanishing $g^S_{RR}$ when the superpartners of LH and RH charged leptons mix\cite{Profumo:2006yu}. The latter effects are too small to generate an observable deviation of $\delta$ from the SM value, since this effect arises at one-loop level and since $g^S_{RR}$ enters $\delta$ quadratically. On the other hand, this parameter interferes linearly with $g^V_{LL}$ in the quantity $\eta$, which is assumed to be zero when the value of the Fermi constant is extracted from the muon lifetime, $\tau_\mu$. If the L-R superpartner mixing is large, the correction to $\tau_\mu$ could be large enough to affect the value of $G_\mu$ to be determined from the next generation of muon lifetime experiments currently underway at PSI.

It turns out the scale of the mass of light neutrinos has significant implications for the values of some of the $g^\gamma_{\epsilon\mu}$. In particular, for $\epsilon\mu=LR$ or $RL$ -- corresponding to neutrinos of opposite chirality -- radiative corrections involving the associated operators can generate contributions to the neutrino mass matrix. In the absence of \lq\lq unnatural" large cancellations between these radiative contributions and tree-level neutrino mass terms,  these radiative contributions should  not be large compared to the scale of entries in the neutrino mass matrix. Based on such \lq\lq naturalness" considerations,  one would expect the $|g^\gamma_{RL,LR}|$ to be bounded from above by the scale of $m_\nu$. These implications were recently noted in Ref.~\cite{Prezeau:2004md} and followed up in several studies, considering both weak decays \cite{Erwin:2006uc} and neutrino magnetic moments\cite{Bell:2005kz,Davidson:2005cs,Bell:2006wi}.

In analyzing the neutrino mass naturalness implications, one has to be careful to employ a set of gauge invariant operators  [the four fermion operators  in Eq.~(\ref{eq:leff0}) are not] while taking into account the flavor structure of all gauge invariant operators that can contribute to muon decay\cite{Erwin:2006uc}. As a result, one finds that the scale of $m_\nu$ implies rigorous bounds on the $|g^V_{LR,LR}|$ ($\lsim 10^{-4}$) and constrains some, but not all, of the gauge invariant four fermion operators that can give rise to scalar and tensor interactions involving neutrinos of opposite chirality.
Thus, if future measurements of these PV Michel parameters yielded non-zero deviations from the SM, one would likely conclude that the source is one of the gauge invariant four fermion operators
\begin{equation}
\epsilon^{ij}\bar{L}_{i}^A\ell_{R}^C\bar{L}_{j}^B\nu_{R}^D
\end{equation}
with generation indices $A=B=1$, $C=2$ or $A=B=2$, $C=1$, as these operators cannot contribute to the neutrino mass matrix through radiative corrections\footnote{Here, $L$ is an SU(2)$_L$ doublet, $\ell_R$ is an SU(2)$_L$ charged lepton singlet, and $i,j$ are generation indices.}. 

An analogous situation pertains in the semileptonic sector. For weak decays of systems containing light quarks, one can write down an effective Lagrangian analogous to Eq.~(\ref{eq:leff0}):
\begin{equation}
\label{eq:leffbeta}
{\cal L}^{\beta-\rm decay} = - \frac{4 G_\mu}{\sqrt{2}}\ \sum_{\gamma,\, \epsilon,\, \delta} \ a^\gamma_{\epsilon\delta}\, 
\ {\bar e}_\epsilon \Gamma^\gamma \nu_e\, {\bar u} \Gamma_\gamma d_\delta
\end{equation}
where the $a^\gamma_{\epsilon\delta}$ play the role of the $g^\gamma_{\epsilon\mu}$ in this sector. There exist several equivalent representations of the low-energy effective semileptonic interaction\cite{Severijns:2006dr,Herczeg:2001vk}, but I prefer the form in Eq.~(\ref{eq:leffbeta}) because of its similarity to the muon decay effective Lagrangian. In the Standard Model, one has $a^V_{LL}=V_{ud}$ at tree level. Beyond tree-level, one must correct this expression for the difference between radiative corrections to the muon and $\beta$-decay amplitudes because we have normalized ${\cal L}^{\beta-\rm decay}$ in terms of the muon decay Fermi constant. 

As members of this community well know, obtaining a reliable determination of $a^V_{LL}$ is of considerable interest since the value of $V_{ud}$ is needed test the unitarity of the CKM matrix. This subject has been discussed extensively elsewhere\cite{Ramsey-Musolf:2006vr,Severijns:2006dr,Blucher:2005dc}, so I will not dwell on the details here. It is worth noting, however, that in addition to the using of super allowed Fermi transitions nuclei to determine $a^V_{LL}$, considerable effort is being devoted to obtaining this parameter from a combination of the neutron lifetime ($\tau_n$) and one or more of its PV decay correlations that appear in the partial rate\cite{Jackson}:
\begin{eqnarray}
\label{eq:betacor}
   d\Gamma& \propto & {\cal N}(E_e) d\Omega_e d\Omega_\nu d E_e\, \Biggl\{ 1+a {{\vec p}_e\cdot{\vec p}_\nu\over E_e E_\nu}
   + b{\Gamma m_e\over E_e}\\
 \nonumber
    &+& \langle {\vec J}\rangle\cdot \left[A{{\vec p}_e\over E_e} 
   + B{{\vec p}_\nu \over E_\nu} + D{{\vec p}_e\times {\vec p}_\nu \over E_e E_\nu}\right] \\
 \nonumber
&+& {\vec\sigma}\cdot\left[N \langle{\vec J}\rangle + G\frac{{\vec p}_e}{E_e}+Q^\prime {\hat p}_e {\hat p}_e\cdot \langle{\vec J}\rangle+R \langle {\vec J}\rangle\times\frac{{\vec p}_e}{E_e}\right]
 \Biggr\}
  ,
\end{eqnarray}
where ${\cal N}(E_e)=p_e E_e(E_0-E_e)^2$; $E_e$ ($E_\nu$), ${\vec p}_e$ 
(${\vec p}_\nu$), and ${\vec\sigma}$ are the $\beta$ (neutrino) energy, momentum, and polarization, respectively; ${\vec J}$ is the  polarization of the decaying nucleus; and $\Gamma=\sqrt{1-(Z\alpha)^2}$. Since the strong interaction renormalizes the vector and axial vector components of the $V-A$ quark currents in Eq.~(\ref{eq:leffbeta}), and since the neutron decay amplitude depends on both components, one requires both $\tau_n$ and the ratio of  axial vector to vector hadronic couplings, $\lambda=g_A/g_V$. The latter can be obtained, for example, from the correlation coefficients $a$, $A$, or $B$:
\begin{equation}
\label{eq:corcoeff}
a = {1-\lambda^2\over 1+3\lambda^2}, \ \
A = -2{\lambda(1+\lambda)\over 1+3\lambda^2}, \ \
B =  2{\lambda(\lambda-1)\over 1+3\lambda^2}\ ,
\end{equation}
where both $A$ and $B$ are associated with PV correlations. To date, the most precise determinations of the $A$ parameter have been obtained from neutron decay experiments at ILL, while a new measurement of $A$ is underway using ultracold neutrons at LANSCE. Future neutron decay experiments at the SNS will a comprehensive set of correlation measurements that could lead to significantly lower systematic uncertainties in the value of $\lambda$ that achievable with any single measurement alone. 

Measurements of these correlation coefficients can also probe non-$(V-A)\times(V-A)$ interactions that enter Eq.~(\ref{eq:leffbeta}). For example, supersymmetric box graph corrections can generate the scalar and tensor  interactions parameterized by $a^S_{RR}$, $a^S_{RL}$, and $a^T_{RL}$ in the presence of mixing between LH and RH scalar fermions\cite{Profumo:2006yu}. The effect of such interactions would show up most strongly in terms in Eq.~(\ref{eq:betacor}) that depend on $m_e/E_e$, such as the \lq\lq Fierz interference" term proportional to $b$ as well as the PV neutrino correlation parameter, $B$\cite{Ramsey-Musolf:2006vr}: 
\begin{eqnarray}
\nonumber
B^{\rm SUSY}_{\rm box}  =  -2\left(\frac{\Gamma m_e}{E_e}\right)\, \frac{\lambda}{1+3\lambda^2}\,
{\rm Re}\, \Biggl\{ 4\lambda \left(\frac{g_T}{g_A}\right)\, \left(\frac{a^{T}_{RL}}{a^{V}_{LL}}\right)^\ast\\
\nonumber
 +\left[2 \left(\frac{g_T}{g_A}\right) \left(\frac{a^{T}_{RL}}{a^{V}_{LL}}\right)^\ast - \left(\frac{g_S}{g_V}\right)
\left(\frac{a^{S}_{RL}+a^S_{RR}}{a^{V}_{LL}}\right)^\ast\right]\Biggr\}
\end{eqnarray}
where $g_S$ and $g_T$ are scalar and tensor form factors. If the L-R superpartner mixing is near maximal and the superpartner masses are not too heavy, then it is possible that $|B|$ can be as large as $\sim 10^{-3}$, while future correlation experiments could probe the energy-dependent part of $B$ at the few $\times 10^{-4}$ level. Observation of such large effects could be problematic for supersymmetry, as it would imply either fine tuning in order to obtain proper electroweak symmetry-breaking and/or superheavy Higgs scalars that could not be observed at the LHC\footnote{There would still be one light SM-like Higgs boson.}. 

\subsection{PV Electron Scattering}

Future measurements of PV electron scattering asymmetries at Jefferson Lab will provide similarly interesting probes of new physics. For elastic scattering from a target $f$ the asymmetry has the general form\cite{Musolf:1993tb}
\begin{equation}
A_{PV}=\frac{G_{\mu}}{4 \sqrt{2}\pi\alpha}Q^2\left[
Q_W^f+F(\theta, Q^2)
\right],
\end{equation}
where $F(\theta, Q^2)$ is a form factor term and where the weak charge $Q_W^f$ is given by
\begin{eqnarray}
\label{eq:C1f-radcorr}
Q_W^f &= &\hat\rho_{NC}(0) \left[2 I_3^f -4
Q_f\hat\kappa(0,\mu)\sinhat(\mu)\right]\\
\nonumber 
& +& \hat\lambda^f_V +\left(-1+4\hat{s}^2\right)\hat\lambda^e_A+{\rm box}\ \ \ .
\end{eqnarray}
Here, $I_3^f$ is the third component of the target's weak isospin, $\sinhat\equiv\sin^2\hat\theta_W(\mu)$ gives the square of the sine of the weak mixing angle in the $\overline{MS}$ scheme, ${\hat\rho}_{NC}$ and ${\hat\kappa}$ denote a universal set of radiative corrections, $\hat\lambda_{V,A}^f$ are the vertex plus external leg corrections, and \lq\lq $+{\rm box}$" indicate box graph corrections. In the case of SUSY radiative corrections to the weak charges of the electron and proton -- targets of interest to the future Jefferson Lab program -- the dominant effects enter through ${\hat\kappa}$\cite{Kurylov:2003zh}. For the $Q_W^p$, these corrections could be as large as $\sim 4\%$ while for $Q_W^e$ the relative effect could be twice as large. In both cases, the effect of the SUSY corrections always decreases the effective weak mixing angle, given by 
\begin{equation}
\sin^2\hat\theta(Q^2)^{\rm eff}=\hat\kappa(Q^2,\mu)\sinhat(\mu)\ \ \ ,
\end{equation}
thereby leading to a relative {\em increase} in the magnitude of the weak charges. 

It is worth noting that the central value for $\sin^2\hat\theta(0)^{\rm eff}$ obtained from the SLAC E158 M\o ller experiment \cite{Anthony:2005pm} -- though in agreement with the SM prediction at better than the $2\sigma$ level -- is slightly larger than the SM prediction. If the results of the Q-Weak measurement or future Jefferson Lab M\o ller experiment were to agree with this central value but with smaller error bars, the largest SUSY corrections would be disfavored. This could be particularly interesting in light of the value for the muon anomalous magnetic moment, which favors the same SUSY parameter space that would give the largest increases in the weak charges. Thus, one could imagine a situation where the results for $g_\mu-2$ and the PVES measurements could lead to some tension within SUSY.

One way such a situation might be avoided is to allow for so-called \lq\lq R parity violating" (RPV) effects in SUSY\cite{Barger:1989rk}. These interactions are entirely supersymmetric but entail the violation of lepton (L) or baryon number (B). The L-violating interactions are particularly interesting for PVES, where they generate tree-level effects. In the case of the weak charge of the electron, for example, one has \cite{Ramsey-Musolf:1999qk}
\begin{equation}
\label{eq:rpv-weak}
\frac{\delta Q_W^{e}}{Q_W^{e}} \approx -45\, \left(\frac{100\, {\rm GeV}}{\tilde m}\right)^2\, \vert\lambda_{12k}\vert^2
\end{equation}
where $\tilde m$ is the mass of the exchanged scalar lepton and $\lambda_{12k}$ is the relevant L-violating coupling. Note that this effect does not here directly into the PV $ee$ amplitude, but rather indirectly through the normalization of the amplitude to the muon decay Fermi constant. The quantity $\lambda_{12k}$ is, in fact, an RPV coupling that enters that process. 

For $\tilde m=1$ TeV, a 5\% measurement of $Q_W^e$ would be sensitive to $\lambda_{12k}\sim 0.3$. This sensitivity is within a factor of two what has been achieved to date in direct searches for L-violation with $\mu\to e\gamma$ experiments -- a rather remarkable statement of the power for PVES as a probe of new physics. It is also interesting to observe that the effect of RPV interactions is always to {\em decrease} the magnitude of $Q_W^e$ from its SM value, in contrast to the situatio with SUSY loop effects. The E158 results hint in this direction, and the results of a future experiment with substantially smaller error bars could provide stronger indications for RPV effects. Such a result could be significant, because the presence of RPV implies that superpartners can ultimately decay to SM particles, so the lightest supersymmetric particle could not have lived long enough to form the cold dark matter of the universe. In addition, loop effects with the RPV interactions generate a Majorana mass for the neutrino. Thus,  if one had conclusive evidence for RPV in SUSY, one would know that neutrinos are Majorana particles even if future $0\nu\beta\beta$ experiments yielded null results. 

Measurements of the weak charges can provide similarly interesting and complementary probes of other BSM physics scenarios, such as the presence of new light $Z^\prime$ bosons, doubly charged Higgs scalars, leptoquarks, {\em etc.} (for recent discussions, see Refs.~\cite{Erler:2003yk,Czarnecki:2000ic} ).

\section{PV: Electroweak Symmetry-breaking and the Origin of Matter}

An important task for nuclear physics is to explain why there exists any baryonic matter in the universe at all. If, in fact, the universe was matter-antimatter symmetric at the end of the inflationary epoch, the there would have to have been some dynamics in the particle physics of the evolving cosmos to create a matter-antimatter asymmetry. Four decades ago, Sakharov  enumerated the necessary ingredients in those dynamics\cite{Sakharov:1967dj}: (1) a violation of baryon number; (2) the presence of both C and CP-violation; and (3) a departure from thermal equilibrium at some juncture\footnote{The latter assumes that CPT is an exact symmetry. If CPT was violated, then a baryon asymmetry could have been created during equilibrium dynamics.}. In principle, the SM contains all three ingredients. Baryon number violation arises through anomalous processes called \lq\lq sphaleron transitions" between different electroweak vacua. The C-violation needed arises through the axial vector couplings of gauge-bosons to fermions, while the CP-violation appears via the complex phase in the CKM matrix. Finally, when the universe cooled through the electroweak temperature and the Higgs got its vacuum expectation value, one could have seen a departure from thermal equilibrium. 
It is well known, however, that neither the CP violation implied by observations in the kaon and B-meson systems, nor the electroweak phase transition (EWPT) in the SM, are sufficiently strong to explain the observed baryon asymmetry, characterized here by the baryon to photon entropy density ratio\footnote{The mass of the Higgs boson is too heavy to allow for a strong first order EWPT in the SM.}:
\begin{eqnarray}
\label{eq:ewb1}
Y_B\equiv \frac{n_B}{s} = 
\biggl\{
\begin{array}{cc}
(7.3\pm 2.5)\times 10^{-11}, & {\rm BBN }\\
(9.2\pm 1.1)\times 10^{-11}, & {\rm WMAP}
\end{array}
\end{eqnarray}
where \lq\lq BBN" and \lq\lq WMAP" indicate values derived from Big Bang Nucleosynthesis\cite{Eidelman:2004wy} and the cosmic microwave background \cite{Spergel:2003cb}, respectively. 

Clearly, it is up to BSM physics to account for  the small, but anthropically crucial, value of $Y_B$. A number of BSM scenarios that may provide such an explanation have been proposed and discussed extensively in the literature. Here, I wish to focus on the possibility that new physics during the era of electroweak symmetry-breaking was responsible. In SUSY, for example, there exist abundant sources of new CP violation whose effects are not {\em a priori} suppressed as in the SM, and the SUSY Higgs sector may provide for the requisite strong first order EWPT. It is timely to think about these and other electroweak scenarios because both the LHC and precision measurements will be probing physics at this scale. If nothing else, we have some chance of testing and either ruling out electroweak scale baryogenesis (EWB) or providing evidence for its viability. In this respect, PV will play a key role.

The most powerful probes of new CP-violation present during the EWPT are searches for the permanent electric dipole moments of various elementary particles, atoms, and nuclei. As discussed elsewhere\cite{Erler:2004cx,Ramsey-Musolf:2006vr,Pospelov:2005pr}, we are on the brink of a revolution in EDM searches, as experimentalists expect to push the sensitivity of these measurements to BSM CP-violation by several orders of magnitude during the LHC era. These measurements will not be sensitive enough to observe EDMs predicted in the electroweak sector of the SM, but they could uncover the effects of BSM CP violation. Thus, if new electroweak scale physics is discovered at the LHC; if LHC and future linear collider studies of the Higgs sector show that a strong first order EWPT is viable with such new physics; and if the EDM searches yield non-zero results, the possibility that the baryon asymmetry was produced at the electroweak symmetry-breaking era would be quite compelling. 

The basic idea of an EDM search is quite simple. One peforms a Larmour precession measurement with a sample of, {\em e.g.}, polarized neutrons in a configuration of magnetic and electric fields. The component of the precision frequency due to the EDM, $d{\vec J}$, is given by
\begin{equation}
\nu^{\rm EDM} = -\frac{d{\vec J}\cdot{\vec E}}{h}\ \ \ .
\end{equation}
Note that this effect is odd under both time reversal and parity. In contrast, the Larmour frequency due to the interaction of the magnetic field and magnetic moment are P and T-even. To separate the two effects, one exploits the PV aspect of $\nu^{EDM}$ and looks for a change in the precession frequency upon reversal of the direction of $\vec E$. This reversal amounts to performing a parity transformation since $\vec E$ changes sign under such a transformation but $\vec J$ does not. 

Searches for the EDM of the electron, neutron, muon, and neutral atoms have been pursued for many years, with increasingly stringent upper bounds on the magnitudes of EDMs being achieved in each case. For recent reviews of EDM searches, see, {\em e.g.}, Refs.~\cite{Erler:2004cx,Ramsey-Musolf:2006vr,Pospelov:2005pr}. These null results can have significant implications for the viability of EWB. In a simple supergravity scenario in supersymmetric baryogenesis, $Y_B$ can depend on two phases\cite{Huet:1995sh,Carena:2002ss,Lee:2004we,Konstandin:2005cd}:
\begin{equation}
\label{eq:yb}
Y_B\equiv\frac{\rho_B}{s_\gamma} = F_1\, \sin\phi_\mu + F_2\, \sin(\phi_\mu+\phi_A)\ \ \ ,
\end{equation}
where $\phi_\mu$ and $\phi_A$ are the CP-violating phases associated with the supersymmetric \lq\lq $\mu$ term and SUSY-breaking tri-scalar \lq\lq $A$" terms, respectively. The coefficients $F_{1,2}$ depend on the other parameters of the SUSY model and on the detailed transport dynamics during the electroweak phase transition. The latter entails a detailed competition between CP-violating asymmetries in the scattering of superpartners from the spacetime varying Higgs vacuum expectation values and CP-conserving interactions that like to minimize free energy by causing these asymmetries to relax to zero. 

It turns out that for this scenario, the most important term is the $F_1$-term. To obtain the observed baryon asymmetry, the corresponding value of $\sin\phi_\mu$ has to be greater in magnitude than about $0.2$. Present limits on the EDM of the electron, for example, imply that $|\sin\phi_\mu|$ is considerably smaller, if the mass of the selectron is below about a TeV. In this case, $d_e$ is dominated by its one-loop contribution and this scenario for EWB is close to being ruled out. On the other hand, if the selectron mass is larger than a few TeV, the $d_e$ becomes two-loop dominated and larger CP-violating phases are  -- consistent with EWB -- are allowed by present EDM limits. Taking the minimal value for $|\sin\phi_\mu|$ allowed by the observed baryon asymmetry and the two loop EDM, one would expect a value for $|d_e|$ of order $10^{-28}$ $e$-cm or larger if this scenario for EWB is to remain viable\cite{Cirigliano:2006dg}. A similar conclusion applies to the neutron EDM. New experiments being carried out for both the electron and neutron could reach this level of sensitivity within the next several years -- making the coming period particularly interesting for the interplay of EDMs and cosmology. 

\section{Conclusions}

It seems clear to me that the coming decade will be a period of intense interest in the studies of PV interactions at low- and intermediate-energies. The field has come a long way from its inception 50 years ago and blossomed into a remarkably rich and diverse area of physics. Recent experimental advances, together with new theoretical developments, have put the field on the cusp of a new era. Refining this tool to probe both BSM physics as well as the structure and dynamics of hadrons should engage our efforts for many years to come.

\vskip 0.1in

\noindent{\bf Acknowledgments} I would like to thank B. Holstein, B. Desplanques, R. McKeown, K. Kumar, P. Souder, C.-P. Liu, and C. Keppel for useful conversations. This work was supported under U.S. Department of Energy contract DE-FG02-05ER41361 and National Science Foundation Award PHY-05556741. 

%
%
%
%
%

\end{document}